\documentclass[prb,preprint]{revtex4-1} 

\usepackage{graphicx} 
\usepackage{epstopdf}
\newcommand{\e}{{\rm e}}
\newcommand{\dr}{{\Delta r}}

\newcommand{\bea}{\begin{eqnarray}}
\newcommand{\eea}{\end{eqnarray}}
\newcommand{\be}{\begin{equation}}
\newcommand{\ee}{\end{equation}}
\newcommand{\ba}{\begin{eqnarray}}
\newcommand{\ea}{\end{eqnarray}}

\newcommand{\ra}{\rightarrow}

\newcommand{\la}{\label}

\begin{document}


\title{The hardwall method of solving the radial Schr\"odinger equation and 
unmasking hidden symmetries}

\author{Siu A. Chin}
\email{chin@physics.tamu.edu} 
\author{John Massey}
\email{john.massey@tamu.edu}
\affiliation{Department of Physics and Astronomy, Texas A\&M University, College Station, Texas 77843-4242}



\date{\today}

\begin{abstract}
Solving for the bound state eigenvalues of the Schr\"odinger equation is a tedious iterative
process when the conventional shooting or matching method is used.
In this work, we bypass the eigenvalue's dependence on the eigenfunction by simply trying out all 
eigenvalues to a desired accuracy. When the eigenvalue is known, the integration for the eigenfunction 
is then trivial. At a given energy, by outputting the radial distance at which the wave function crosses zero (the hardwall radius), this method automatically determines the entire spectrum of eigenvalues of the radial Schr\"odinger equation without iterative adjustments. 
Moreover, such a spherically symmetric ``hardwall'' can unmask ``accidental degeneracy''
of eigenvalues due to hidden symmetries. We illustrate the method on the Coulomb, harmonic, Coulomb+harmonic, and the Woods-Saxon potentials.  

\end{abstract}

\maketitle 

\section{Introduction} 

Solving the radial Schr\"odinger equation for a central potential $V(r)$,
\begin{equation}
-\frac{\hbar^{2}}{2m}\left(\frac{\partial^{2}}{\partial r^2} 
- \frac{l(l+1)}{r^{2}}\right)u(r) + V(r)u(r) = E u(r),
\label{re1}
\end{equation}
where $u(r)=r R_{nl}(r)$, has been considered
an important teaching tool in this Journal\cite{bol} and a standard subject in
Koonin's early computational physics text.\cite{koo}
However, in recent years, this topic is not discussed in most computational physics texts,\cite{gou,gio,dev,ves,lan}
or only alluded to briefly.\cite{gio} Perhaps, this is related to the difficulty of
implementing the conventional  shooting or matching method \cite{bol,gou,gio,dev,ves,lan} 
of solving the eigenvalue problem. While more sophisticated methods\cite{ixa,GVB WS,sac,aug}
have been suggested in the literature, we introduce here a classroom-tested, start-from-scratch method, 
that must be the simplest among all known ways of solving the bound state problem of the radial Schr\"odinger equation.

Our simplification follows from two key ideas: 1) The eigenvalue $E$ is
usually determined by requiring the eigenfunction to satisfy
certain boundary conditions. The adjustment of the eigenvalue by 
shooting or matching the eigenfunction\cite{gou,gio,dev,ves,lan} to
satisfy the boundary condition is a tedious iterative process. In this work, we completely 
by-pass this adjustment process by sweeping through all values of $E$ to 
a desire accuracy. That is, we simply try all values of $E$, up to a certain precision, 
by brute-force. When $E$ is known, the eigenfunction can be obtained easily. 
This then breaks the dependence of the eigenvalue to the eigenfunction.
2) To decide which $E$ is the correct eigenvalue, we observe that
{\it every} $E$ value is an eigenvalue of the given potential {\it plus} 
an infinite potential barrier at {\it some} radius  $C$. The correct eigenvalue $E$ is 
then obtained in the limit of $C\ra\infty$. Given $E$, $C$ is the simply the
radial distance {\it wherever} the wave function vanishes. This is then the ``hardwall''
condition to be described in Section {\ref{hwc}}. Our method takes advantage of the
fact that it is far easier to find $C$ as a function of $E$, than the other way around,
as in the matrix method.\cite{aug} Also in contrast to the matrix method, 
no basis functions need to be assumed, no matrix elements
need to computed and no matrix needs to be diagonalized.

This brute-force approach is conceptually 
very simple and can be easily understood by any undergraduate taking a first course 
in quantum mechanics. However, this approach, by leveraging the power of modern computers, 
can decipher the entire spectrum of a central potential automatically without 
piecemeal adjustments. Moreover, this method can unmask hidden symmetries
by revealing ``accidental degeneracy'' of eigenvalues and their patterns of
degeneracy.
 
\section{Solving the radial Schr\"odinger equation numerically}

To numerically solve the reduced radial Schr\"odinger equation (\ref{re1}),
we will first cast it into a  dimensionless form in terms of units most 
appropriate for the equation. This digression seems appropriate 
since most current texts\cite{gou,gio,dev,ves,lan} do not emphasize this important point.

To arrive at a dimensionless equation, one first sets 
$r=r^*a$, where $a$ is a unit of length to be determined, 
and $r^*$ is just a dimensionless number. The equation then reads
$$
-\frac{\hbar^{2}}{2ma^2}\left(\frac{\partial^{2}}{\partial r^{*2}} - \frac{l(l+1)}{r^{*2}}\right)u(r^*) + V(r^* a)u(r^*) = E u(r^*).
$$
Since $\e_0\equiv\hbar^2/ma^2$ is a unit of energy, it is natural to write $E=E^*e_0$, so that
(\ref{re1}) has the dimensionless form
\be
-\frac{1}{2}\left(\frac{\partial^{2}}{\partial r^{*2}} - \frac{l(l+1)}{r^{*2}}\right)u(r^*) + v(r^*)u(r^*) = E^* u(r^*),
\la{se2}
\ee
where the dimensionless potential is given by
$$
v(r^*)=\frac{V(r^* a)}{e_0}=\frac{ma^2V(r^* a)}{\hbar^2}.
$$
One then chooses $a$ so that $v(r^*)$ is as simple as possible. For example, in the case of the
Coulomb potential with $V(r)=-ke^2/r$, the dimensionless potential is
$$
v(r^*)=-\frac{ma^2 ke^2}{r^*a\hbar^2}=-\frac{1}{r^*},
$$
when one chooses $a=\hbar^2/(mke^2)$, which is the Bohr radius. 
The unit of energy $\e_0$ is then the Hartree (Ha) and the spectrum of the hydrogen 
atom is $E=-1/(2n^2)$ Ha. (Note that 1 Ha=2 Ry (Rydberg)=2(13.6 eV). ) This choice
defines the atomic units, most appropriate for dealing with atomic problems. Another example is the
the harmonic potential, with $V(r)=(1/2)m\omega^2 r^2$. In this case
$$
v(r^*)=\frac{ma^2 m\omega^2a^2r^{*2}}{2\hbar^2}=\frac{1}2 {r^{*2}},
$$
with the natural choice of $a=\sqrt{\hbar/m\omega}$, which is the harmonic length, and
$\e_0=\hbar^2/m a^2=\hbar\omega$. These are then the harmonic units.

The advantage of solving a dimensionless equation is that one solves an entire {\it class} of equations at once, not just one particular equation. For example, if one is interested in solving the Coulomb potential with a
nuclear charge $Ze$, then there is no need for a new calculation. One simply changes the units from
$k\ra k Z$. One then sees immediately that the Bohr radius will shrink by a factor of $Z$ and the
energy grow by a factor of $Z^2$. If one is interested in the case of muonic atoms, where the
electron is replaced by the muon 207 times as massive, then the radius of muonic-hydrogen will
be 207 times smaller with binding energy $(207)^2$ times greater.
Similarly for the harmonic case; the dimensionless equation
solves an entire class of problems with any values of $m$ and $\omega$. 
These important insights are lost when explicit units are used, as done 
in some computational texts.\cite{dev}

From this point forward, we will drop all asterisks $^*$ in referencing the dimensionless radial equation (\ref{se2}).

The eigenvalue problem associated with the radial equation (\ref{se2}) 
is that one must determine $E$
subject to the boundary conditions $u(0)=0$ and $u(\infty)=0$. 
The shooting method starts out at $u(0)=0$, integrates out to some
large value of $r=R$, and adjusts $E$ so that $u(R)=0$. (The needed
$R$ value is different for different eigenstates.)

In this work, as in the shooting method, we will also start at the origin with $u(0)=0$
and integrate outward, but will not need to impose the condition $u(R)=0$ for 
an {\it a priori} unknown $R$ value. 
Instead, we will introduce the hardwall condition to be described in the next Section.

The dimensionless radial equation (\ref{se2}) can be further arranged as
\begin{equation}
\frac{\partial^{2}u(r)}{\partial r^{2}} = f(r)u(r),
\label{sch} 
\end{equation}
where
$$
f(r)=\frac{l(l+1)}{r^{2}} +2V(r) - 2E. 
$$
It is well-known that (\ref{sch}) can be efficiently solved by use of the fourth-order Numerov\cite{num} algorithm, which has been derived numerous time in this Journal,\cite{cho,bol} and elsewhere:\cite{Hartree, koo}
$$
(1-\frac{1}{12}\dr^2f_{n+1})u_{n+1}=(2+\frac56 \dr^2 f_n)u_n-(1-\frac{1}{12}\dr^2f_{n-1})u_{n-1},
$$
where $u_n=u(n\dr)$ and $f_n=f(n\dr)$. 
For a given $\dr$, one can define  $g_n=1-\frac{1}{12}\dr^2f_{n}$ so that the algorithm reads simply,
\be
u_{n+1}=\frac{(12-10g_n)u_n-g_{n-1}u_{n-1}}{g_{n+1}}.
\la{numerv}
\ee
The above Numerov algorithm is used to obtain
all results reported in this work.

\section{The hardwall condition}
\la{hwc}

The hardwall method replaces the $u(\infty)=0$ boundary condition with the following hardwall
condition: for a set of given $l$ and $E$ values, $u(r)$ is iterated outward using (\ref{numerv})
at a chosen $\dr$, with $u_0=0$ and $u_1$ an arbitrary but small real number,
out to a large value of $r=R$, depending on the number of eigenvalues desired. 
Whenever $u_n$ crosses zero, {\it i.e}, $u_{n-1}u_{n}<0$, 
we output $(C,E)$ where $E$ is the given energy and $C=(n-1/2)\dr$ is the hardwall radius.
That is, whenever $u_n$ crosses zero, $E$ is the {\it  exact} energy corresponding to 
the original potential $v(r)$ plus an infinite potential (the ``hardwall'') at $r=C$. 
In other words, every value of $E$ is an eigenvalue for $v(r)$ plus a hardwall 
potential at {\it some} values of $C$. For each value of $l$, one does the above outward
integration of $u_n$ over a set of prescribed $E$ values spanning the bound state range of $v(r)$.
When $E$ versus $C$ is plotted after one has swept through an interval of $E$,
one will see that $E$ converges to the eigenvalue of $v(r)$ in the limit of large $C$. 

\begin{figure}
	\centering
	\includegraphics[width=0.90\linewidth]{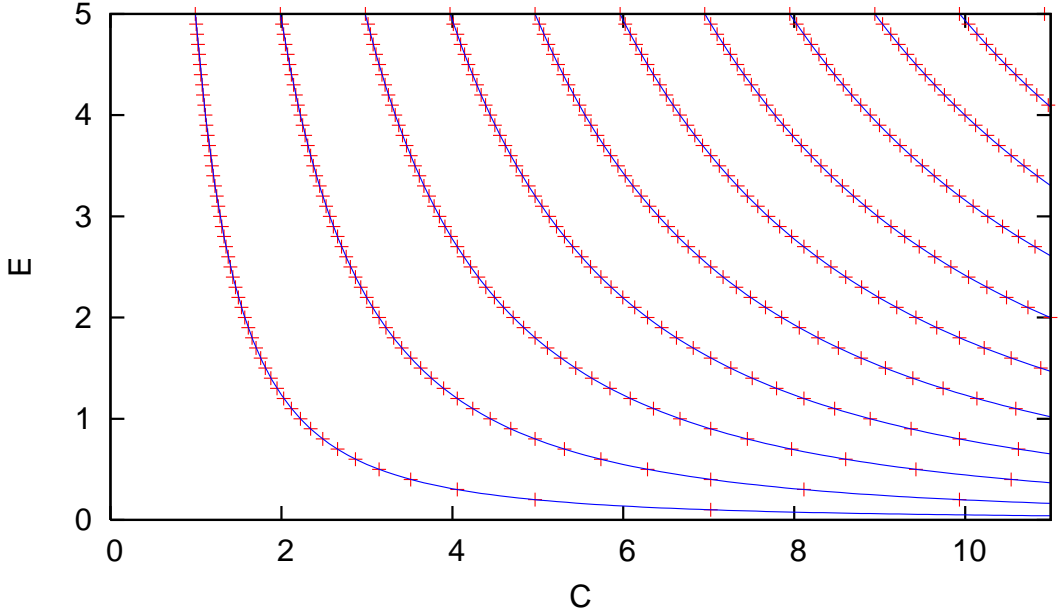}
	\caption{Symbols are energy eigenvalues using the hardwall method for $v=0$. The first ten eigenvalues for $l=0$ are shown. Solid lines are the exact energies $E=(1/2)(n\pi/C)^2$ for $n=1$ to 10.}
	\label{vzero}
\end{figure}

Let's see how the hardwall method works for the null-potential case of $v(r)=0$. 
In this case, the solutions to the radial Schr\"odinger equation 
are spherical Bessel functions $j_{l}(k_{ln} r)$, with energy eigenvalues $E_{ln}=(1/2)k_{ln}^{2}.$ For $l=0$, $j_{0}=\sin(k_{0n}r)/k_{0n}r$. For an infinite wall at the hardwall radius $C$, the wave function must vanish, forcing $k_{0n}C=n\pi$ and $E=(1/2)(n\pi/C)^{2}.$ In Fig.\ref{vzero}, we compare the 
outputted $(C,E)$ of the hardwall method with this analytical result. We shall refer to this $(C,E)$ plot
of the hardwall method as a ``$C$-scan'' of the potential.
For this calculation, we set $l=0$ and iterate over an outer loop
of 50 values of $E_i=i\Delta E$, with $\Delta E=0.1$ and an inner loop of 1100 values of $r=j\dr$ with
$\dr=0.01$ out to $C=11$ (so that only 10 eigenvalues are visible). This calculation demonstrates
 the essential characteristic of the method: 1) At any fixed value of $C$, the $C$-scan gives the
correct eigenvalues of the hardwall potential at $r=C$ when $\Delta E$ is sufficiently fine.
The power of the method resides in the fact at a {\it fixed} $E$, {\it multiple} values of $C$ are 
found simultaneously. If one does the opposite, fixing $C$ then finding $E$, then $E$ must be
determined one by one.
2) As $C\rightarrow\infty$, all eigenvalues approach zero, as they should. 
However, higher eigenvalues approach zero more slowly at larger $C$.
Thus when $v(r)\ne 0$, the $C$-scan will approach the eigenvalues of $v(r)$,
also with higher eigenvalues converging at larger values of $C$.
3) A single sweep in energy produces all the eigenvalues at the same time. 
An accuracy of 5-6 significant digits in determining $E$ is easily achievable on a laptop computer.

We will now see how effective is this method when applied to physical potentials.

\section{Hidden symmetries in the Coulomb and harmonic potentials}

\begin{figure}
		\includegraphics[width=0.90\linewidth]{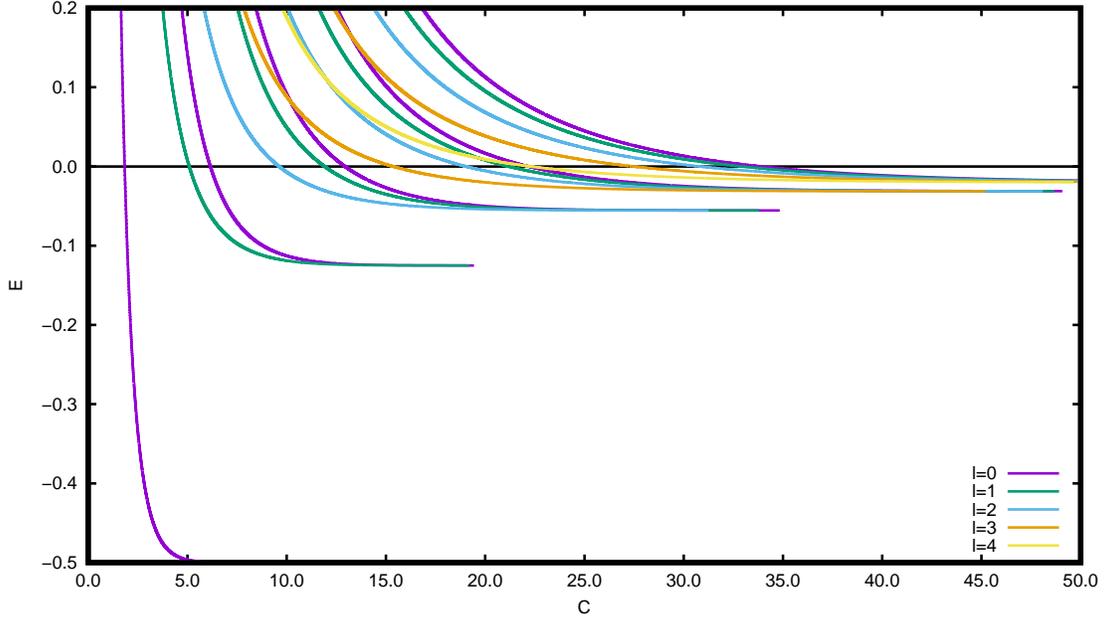} \\
	\caption{The convergence of energy eigenvalues using the hardwall method for the 3D Coulomb potential.}
	\label{coul}
\end{figure}

For the Coulomb potential $v_{\rm C}(r) = -1/r,$ the $C$-scan showing
the convergence of the eigenvalues is shown in
Fig.\ref{coul}. One immediately notices something unusual. In the large
$C$ limit, distinct higher energy levels corresponding to different radial quantum numbers $n_r$ and $l$
coalesce into a single level. Since the hardwall is spherically symmetric, its addition will not
alter the spherical symmetry of the Coulomb potential. For spherical symmetry, each energy level
with quantum number $l$, regardless of $n_r$, is $2l+1$ degenerate, meaning that 
all $2l+1$ states with azimuthal quantum number $m=-l,\cdots, -1,0,1,\cdots, l$ have the same energy. 
However, as the hardwall is gradually removed in letting $C\ra\infty$, further degeneracy is revealed, 
due to the hidden $SO(4)$ symmetry\cite{bay} of the Coulomb potential.

The pattern of this degeneracy is also visible. The
ground state $(l,n_r)=(0,0)$ is non-degenerate. The next two states  $(0,1)$, $(1,0)$
are degenerate. The next three states $(0,2)$, $(1,1)$, $(2,0)$ are degenerate, etc., with the
energy only depending on the sum of $l$ and $n_r$ as $E=-1/(2n^2)$,
with $n=l+n_r+1$. Thus the hardwall method not only computes the eigenvalues,
its $C$-scan also reveal the hidden symmetry and the pattern of degeneracy of the Coulomb potential. 
The resulting eigenvalues are shown in Table \ref{could}.

\begin{table}[h]
	\centering
	\caption{Calculated eigenvalues of the Coulomb potential. The first column is the exact results $-1/(2n^2)$
with $n=l+n_r+1$. }
	\begin{ruledtabular}
		\scalebox{0.8}{
		\begin{tabular}{l c c c c c}
			$-1/(2n^2)$ & $l=0$ & $l=1$ & $l=2$ & $l=3$ & $l=4$ \\
			\hline
			-0.50000 & -0.49998 &  &  &  & \\
			-0.12500 & -0.12499 & -0.12499 &  &  & \\
			-0.05555 & -0.05555 & -0.05555 & -0.05554 &  & \\
			-0.03125 & -0.03124 & -0.03124 & -0.03124 & -0.03123 & \\
			-0.02000 & -0.01984 & -0.01987 & -0.01991 & -0.01996 & -0.01998 \\
		\end{tabular}
	}
	\end{ruledtabular}
	\label{could}
\end{table}

\begin{figure}
\includegraphics[width=0.90\linewidth]{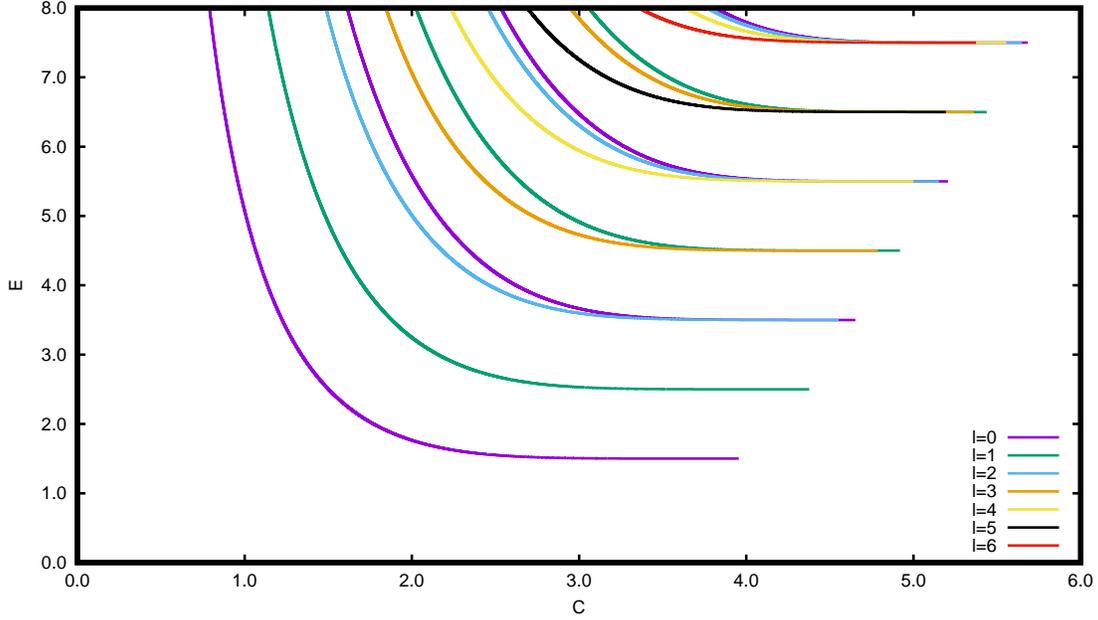} 
	\caption{The convergence of energy eigenvalues for the 3D harmonic potential.}
	\label{hofig}
\end{figure}

For the harmonic potential $v_{\rm H}(r) = (1/2)r^{2},$ the corresponding $C$-scan
is shown in Fig.\ref{hofig}. Again, one observes energy degeneracy 
at large value of $C$. One can therefore conclude that the 3D
harmonic oscillator also has a hidden symmetry (that of $SU3$).\cite{fra} 
Let's see whether one can also deduce its
pattern of degeneracy. There are two states $(l,n_r)=(0,0), (1,0)$ 
that are non-degenerate. There are two states that are twice degenerate:
[(0,1), (2,0)] and [(1,1), (3,0)], thrice degenerate:
[(0,2), (2,1), (4,0)] and [(1,2), (3,1), (5,0)], and four-fold degenerate, etc.
The degeneracy pattern is therefore 
for $l$ even, [(0,0)], [(0,1), (2,0)], [(0,2), (2,1), (4,0)],  [(0,3), (2,2), (4,1), (6,0)] etc..
And 
 for $l$ odd, [(1,0)], [(1,1), (3,0)], [(1,2), (3,1), (5,0)],  [(1,3), (3,2), (5,1), (7,0)] etc..
(Degenerate levels are grouped together by square brackets.)
Both patterns of degeneracy can be accounted for if the energy only depends on
$l+2n_r$. Therefore, one can deduce that the energy spectrum must be given 
by $l+2n_r+3/2$ where $3/2$ is the ground state energy of (0,0). This analytical
result is compared to the computed spectrum given in Table \ref{hod}.

\begin{table}[h]
	\centering
	\caption{Calculated eigenvalues of the harmonic potential. The first column is the exact results $(l+2n_r+3/2)$. }
	\begin{ruledtabular}
		\scalebox{0.9}{
		\begin{tabular}{l c c c c c c c}
			$(l+2n_r+3/2)$ & $l=0$ & $l=1$ & $l=2$ & $l=3$ & $l=4$ & $l=5$ & $l=6$ \\
			\hline
			1.50000 & 1.50002 &  &  &  &  &  &  \\
			2.50000 &  & 2.50001 &  &  &  &  &  \\
			3.50000 & 3.50001 &  & 3.50002 &  &  &  &  \\
			4.50000 &  & 4.50001 &  & 4.50002 &  &  &  \\
			5.50000 & 5.50001 &  & 5.50001 &  & 5.50001 &  & \\
			6.50000 &  & 6.50001 &  & 6.50001 &  & 6.50002 &  \\
			7.50000 & 7.50001 &  & 7.50001 &  & 7.50001 &  & 7.50002 \\
		\end{tabular}
	}
	\end{ruledtabular}
	\label{hod}
\end{table}

\begin{figure}[h]
	\centering
	\includegraphics[width=4.4in,height=3.4in]{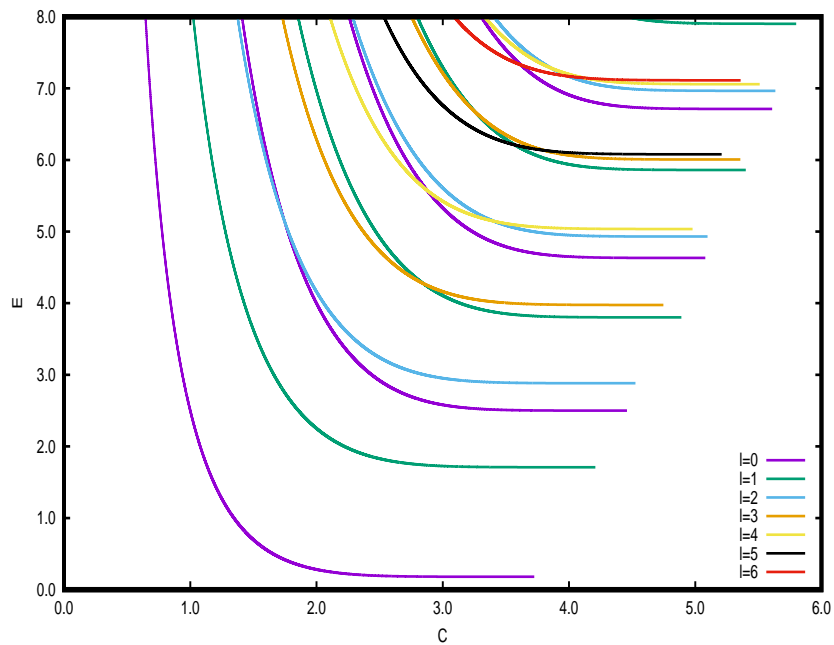}
	\caption{The convergence of energy eigenvalues for the Coulomb + harmonic potential.}
	\label{c+hfig}
\end{figure}

The hidden symmetry of the Coulomb and the harmonic potential is unique in each case. 
If one simply adds the Coulomb and the harmonic potential together, then the hidden symmetry 
is destroyed for the combined potential.
This is illustrated in the $C$-scan of Fig.\ref{c+hfig}. One sees that energy
levels which were $k$-fold degenerate in the harmonic oscillator case, now split
into $k$ energy levels in the large $C$ limit.
The computed eigenvalues are shown in
Table \ref{c+hd}. Since the Coulomb potential is only strong near the origin,
it acts as a perturbation, and only
modifies the low-lying spectrum of the harmonic oscillator.

\begin{table}[h]
	\centering
	\caption{Calculated eigenvalues of the Coulomb+harmonic potential. There is now no degeneracy;}
	\begin{ruledtabular}
		\scalebox{0.8}{
		\begin{tabular}{c c c c c c c c}
			$n_r$&$l=0$ & $l=1$ & $l=2$ & $l=3$ & $l=4$ & $l=5$ & $l=6$ \\
			\hline
			0&0.17968 &  &  &  &  &  &  \\
			1&& 1.70903 &  &  &  &  &  \\
			2&2.50001 &  & 2.88224 &  &  &  &  \\
			3&& 3.80193 &  & 3.97553 &  &  &  \\
			4&4.63196 &  & 4.93068 &  & 5.03608 &  & \\
			5&& 5.86036 &  & 6.00654 &  & 6.07947 &  \\
			6&6.71260 &  & 6.96584 &  & 7.05815 &  & 7.11255 \\
		\end{tabular}
	}
	\end{ruledtabular}
	\label{c+hd}
\end{table}

\section{Woods-Saxon potential}

\begin{figure}[h]
	\centering
	\includegraphics[width=0.90\linewidth]{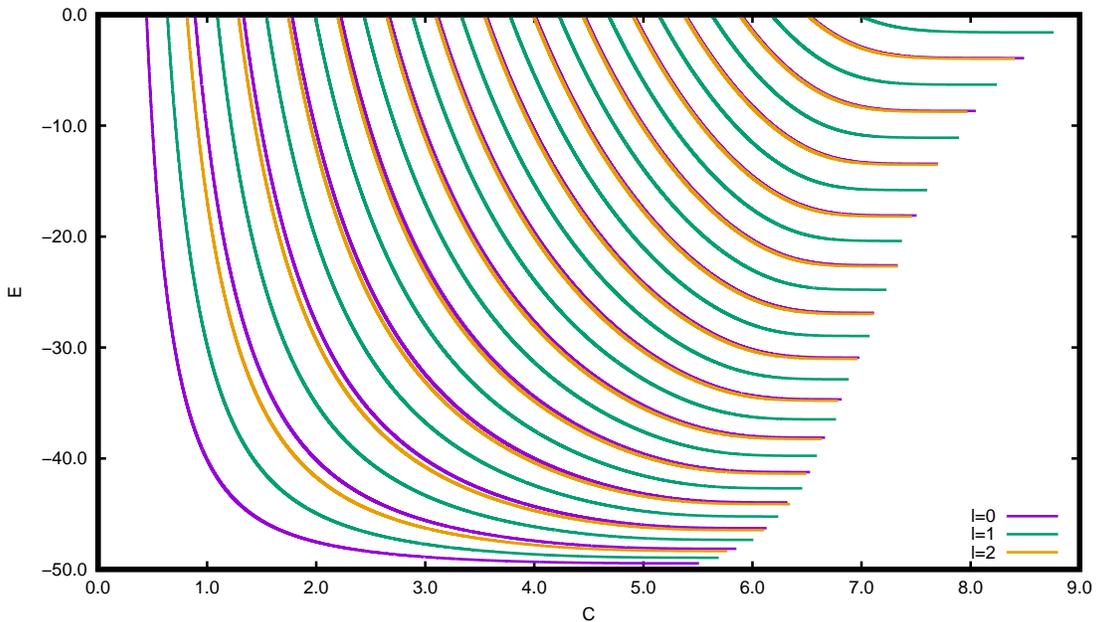}
	\caption{The convergence of energy eigenvalues using the hardwall method for the Woods-Saxon potential
as parameterized by Vanden Berghe\cite{GVB WS}.}
	\label{WS GVB fig}
\end{figure}

The Woods-Saxon potential \cite{WS orig, GVB WS} describes the effective average force confining each nucleon to the interior of the nucleus in the form of 
\begin{equation}
V_{\rm WS}(r)=\frac{u_{0}}{1+t}+\frac{u_{1} t}{(1+t)^{2}},\qquad
t = \exp[\frac{r-r_{0}}{a}],
\end{equation}
with $u_{0}$ and $u_{1}$ fixing the potential well depth, $a$ the surface thickness of the nucleus, and $r_{0}$ the nuclear radius proportional to the mass number. 
We use the parameterization by Vanden Berghe, Fack, and De Meyer\cite{GVB WS} 
with $u_{0}=-50$ MeV, $r_{0}=7$ fm, $a=0.6$ fm, and $u_{1}=-u_{0}/a$. 
In this calculation, we use the dimensionless form of the equation but restate results in
units as given above for comparison.
The convergence of the eigenvalues, with $\dr=0.001$, $\Delta E = 0.0005$,
is shown in Fig.\ref{WS GVB fig}. Since the Woods-Saxon potential has no special symmetry, 
its $C$-scan shows close-by energy levels, but no actual degeneracy.
The converged eigenvalues, a selected set of 21, are compared with exact results in 
Table \ref{GVB Woods-Saxon data}. Our eigenvalues agree with published results up to five significant digits.

\begin{table}[h]
	\centering
	\caption{Calculated eigenvalues as compared to published values \cite{GVB WS} for the Woods-Saxon potential
for angular momentum channels $l=0,1,2$. In each channel, only energies of the even radial quantum state
$n_r$ are compared.  The unit of energy is MeV.}
	\begin{ruledtabular}
		\scalebox{0.8}{
		\begin{tabular}{l c c c c c c}
			 & & Exact results & & & Calculated results & \\
			 $n_r$ & $l=0$ & $l=1$ & $l=2$ & $l=0$ & $l=1$ & $l=2$ \\
			 0 & -49.457788728 & -48.951731623 & -48.34981052 & -49.4570 & -48.9510 & -48.3485 \\
			 2 & -46.290753954 & -45.237176986 & -44.121537377 & -46.2905 & -45.2370 & -44.1215 \\
			 4 & -41.232607772 & -39.767208069 & -38.253426539 & -41.2325 & -39.7670 & -38.2530 \\
			 6 & -34.672313205 & -32.868392986 & -31.026820921 & -34.6720 & -32.8680 & -31.0265 \\ 
			 8 & -26.873448916 & -24.794185466 & -22.689041510 & -26.8730 & -24.7940 & -22.6890 \\
			 10 & -18.094688282 & -15.812724871 & -13.522303352 & -18.0945 & -15.8125 & -13.5220 \\ 
			 12 & -8.676081670 & -6.308097192 & -3.972491432 & -8.6760 & -6.3080 & -3.9720 \\
	 	\end{tabular}
 	}
 	\end{ruledtabular}
 \label{GVB Woods-Saxon data}
 \end{table}

%
%

\section{Conclusion}

In this work, we have demonstrated an extremely simple method of solving the radial 
Schr\"odinger equation by trying out all eigenvalues systematically, up to a certain
precision. The hardwall method relies on the insight that any trial eigenvalue $E$ is an exact eigenvalue
for the given potential plus a hardwall at some radius $C$. By plotting $E$ verses $C$, 
the eigenvalues can be determined to five or more significant digits in
the large $C$ limit. Moreover, this $C$-scan of the spectrum can detect
hidden symmetries by revealing additional degenerate energy levels beyond
spherical symmetry and their patterns of degeneracy.
The method is therefore a valuable tool, not just for solving the radial Schr\"odinger equation,
but also for teaching the emergence of ``accidental degeneracies'' from hidden symmetries.

Finally, the method can also be applied to a strictly 1D Schr\"odinger equation with a symmetric potential $v(x)=v(-x)$. By setting $l=0$, the same starting condition $u_0=0$ now gives the odd-state solutions. 
The even-state solutions can be obtained by starting with $u_0=1$ and $u_1=1$.

\end{document}